\magnification=\magstep1
\overfullrule=0pt
\parskip=6pt
\baselineskip=15pt
\headline={\ifnum\pageno>1 \hss \number\pageno\ \hss \else\hfill \fi}
\pageno=1
\nopagenumbers
\hbadness=1000000
\vbadness=1000000

\vskip 25mm
\vskip 25mm
\vskip 25mm

\centerline{\bf WEYL ORBIT CHARACTERS AND SCHUR FUNCTIONS}
\vskip 5mm

\centerline{\bf H. R. Karadayi}
\centerline{Dept. Physics, Fac. Science, Istanbul Tech. Univ.}
\centerline{80626, Maslak, Istanbul, Turkey }
\centerline{e-mail: karadayi@itu.edu.tr}

\vskip 5mm

\centerline{\bf M. Gungormez}
\centerline{Dept. Physics, Fac. Science, Istanbul Tech. Univ.}
\centerline{80626, Maslak, Istanbul, Turkey }
\centerline{e-mail: gungorm@itu.edu.tr}

\vskip 5mm
\medskip

\centerline{\bf{Abstract}}

For finite Lie algebras, it is shown that characters can be defined first for
Weyl orbits and then for irreducible representations. For $A_N$ Lie algebras,
weight multiplicities can then be calculated by only stating that characters
are equivalent to Schur functions. This also means that to calculate characters
there is no need to sum over Weyl groups. The compatibility with the Weyl
character formula will however be shown.

\vskip 15mm
\vskip 15mm
\medskip

\hfill\eject

\vskip 3mm
\noindent {\bf{I.\ INTRODUCTION}}
\vskip 3mm

We begin by considering Schur functions which are defined {\bf[1]} by
$$ \sum_{M=0}^\infty \ S_M(x_1,x_2,.., x_M) \  z^M \equiv
Exp\sum_{i=1}^\infty \ x_i \ z^i \eqno(I.1) $$
where $ S_0=1 $. This definition includes only some restricted class of Schur
functions $S_M(x_1,x_2,.., x_M)$ which are polinomials of M indeterminates
$x_1,x_2, \dots ,x_M$ and also homogeneous of degree M. Let,
$$ (q_1,q_2, \dots ,q_N) $$
be a partition into N parts. It is called a partition with weight M providing
$$ q_1 + q_2 + \dots + q_N \equiv M  \ \ , \ \
q_1 \geq q_2 \geq ... \geq q_N \geq 0  \ \ .  \eqno(I.2)  $$
Note that there could always be cases for which some of integers
$q_1, \dots q_N$ are chosen to be zero. For any such partition, one can make
a generalization
\vskip 1mm
$$ S_{(q_1,q_2, \dots ,q_N)}(x_1,x_2,..,x_M) \equiv
Det \pmatrix{
S_{q_1-0}&S_{q_1+1}&S_{q_1+2}&\ldots&S_{q_1+M-0}\cr
S_{q_2-1}&S_{q_2+0}&S_{q_2+1}&\ldots&S_{q_2+M-1}\cr
S_{q_3-2}&S_{q_3-1}&S_{q_3+0}&\ldots&S_{q_3+M-2}\cr
\vdots&\vdots&\vdots&\vdots\cr} \eqno(I.3)  $$
\vskip 1mm
\noindent where Det means determinant. Each and every matrix element
$ S_k $ is assumed here to be a function of only k independent parameters
$x_1, x_2, \dots , x_k$ where $ k \leq M $ and $S_{k<0}=0$ .
It is worthwhile to note here that both $ S_M(x_1,x_2,.., x_M) $ and also
$S_{(q_1,q_2, \dots ,q_N)}(x_1,x_2,..,x_M)$ are called Schur functions.

This is, in fact, not the whole story. Some degeneration occurs in the
definition of Schur functions when one wants to identify them with the
characters of irreducible representations of $A_N$ Lie algebras. After
introducing the Weyl orbit characters in the next section, the degeneration
of Schur functions will be briefly explained in section III and the formula
(III.7) will be given. In section IV, we show how one calculates weight
multiplicities by identifying characters directly with Schur functions.
It will be seen that, beside the known ones {\bf[2]}, this gives us a
multiplicity formula for $A_N$ Lie algebras. The compatibility of all these
with the Weyl character formula will be given in the last section.


\vskip 3mm
\noindent {\bf{II.\ CHARACTERS FOR $ A_{N-1} $ WEYL ORBITS}}
\vskip 3mm

Characters are conventionally known to be defined for irreducible
representations. Now we will show that they can be defined first for Weyl
orbits and then for representations. Here, it is essential to use
{\bf fundamental weights} $\mu_I$ (I=1,2,.. N) which are defined,
for $A_{N-1}$ Lie algebras, by
$$ \eqalign {
\mu_1 &\equiv \lambda_1 \cr
\mu_i &\equiv \mu_{i-1}-\alpha_{i-1} \ \ , \ \ i=2,3,.. N. } \eqno(II.1) $$
\noindent or conversely by
$$ \lambda_i = \mu_1 + \mu_2 + .. + \mu_i \ \ , \ \ i=1,2,..N-1. \eqno(II.2) $$
\noindent together with the condition that
$$ \mu_1 + \mu_2 + .. + \mu_N \equiv 0 . \ \ \eqno(II.3)   $$
$\lambda_i$'s and $\alpha_i$'s (i=1,2,..N-1) are {\bf fundamental dominant
weights} and {\bf simple roots} of $A_{N-1}$ Lie algebras. For an excellent
study of Lie algebra technology we refer to the book of Humphreys {\bf[3]}.
We know that there is an irreducible $A_{N-1}$ representation for
each and every dominant weight $\Lambda^+$ which can be expressed by
$$ \Lambda^+ = \sum_{i=1}^N q_i \ \mu_i \ \ , \ \
q_1 \geq q_2 \geq .. \geq q_N \geq 0  \ \ .  \eqno(II.4) $$
\noindent One thus concludes that there is a dominant weight $\Lambda^+$ for
each and every partition $(q_1,q_2, \dots ,q_N)$ with weight
$M=q_1+q_2+ \dots +q_N$. The following definition will then be useful.

\noindent {\bf Definition}:

\noindent Let N, M and k be positive integers such that $M-k \ N \leq M$. Then,
we state $Sub(M \ \lambda_1)$ to be the set of dominant weights corresponding,
via (II.4), to all partitions of $M-k \ N$ for $k=0,1,2, \dots$.

\noindent It is now clear that $Sub(M \ \lambda_1)$
is consist of $M \ \lambda_1$ and all of its sub-dominant weights.
The weight structure of the corresponding irreducible representation
$R(\Lambda^+)$ can then be expressed by the aid of the following
{\bf orbital decomposition}:
$$ R(\Lambda^+) = \sum_{\lambda^+ \in Sub(M \ \lambda_1)} \
m(\lambda^+ \prec \Lambda^+) \ W(\lambda^+) \ \ \eqno(II.5) $$
where $m(\lambda^+ \prec \Lambda^+)$ is the multiplicity of $\lambda^+$
within $R(\Lambda^+)$. Note here that one always has
$m(\lambda^+ \prec \lambda^+) = 1$ and also
$$m(\lambda^+ \prec \Lambda^+) = 0  \eqno(II.6) $$
if $\lambda^+$ is not a sub-dominant weight of $\Lambda^+$. For instance,
the tensor representations with M completely symmetric and antisymmetric
indices correspond, respectively, to two extreme cases which come from the
partitions of M into 1 and M non-zero parts provided that
$$ m(\lambda^+ \prec M \ \lambda_1) = 1 \eqno(II.7)  $$
and
$$ m(\lambda^+ \prec \lambda_M) = 0 \ \ ,
\ \ \lambda^+ \neq \lambda_M  \eqno(II.8) $$
for $\lambda^+ \in Sub(M \ \lambda_1)$. They are, in fact, nothing but
the representations $R(M \ \lambda_1)$ and $R(\lambda_M)$, respectively.
The characters can now be expressed by
$$ ChR(\Lambda^+) = \sum_{\lambda^+ \in Sub(M \ \lambda_1)} \
m(\lambda^+ \prec \Lambda^+) \ ChW(\lambda^+) \ \ \eqno(II.9) $$
In view of orbital decomposition (II.5), it is clear that (II.9) allows us
to define the characters
$$ ChW(\Lambda^+) \equiv \sum_{\mu \in W(\Lambda^+)} \ e(\mu) \eqno(II.10) $$
\noindent for Weyl orbits $W(\Lambda^+)$. \ {\bf Formal exponentials} $e(\mu)$
are defined, for any weight $\mu$, as in the book of Kac {\bf[4]}.
We refer here to a lemma which we previously introduced {\bf[5]}.

\noindent {\bf Lemma}:

\noindent Let $\Lambda^+$ be an $A_{N-1}$ dominant weight which we know to be
identified with a partition $(q_1,q_2, \dots ,q_N)$. Then, any other weight
$ \omega(\Lambda^+) \in W(\Lambda^+) $ is obtained by permutating the
parameters $q_1,q_2,..,q_N$, that is
$$ \omega(\Lambda^+) = q_{i_1} \ \mu_1+q_{i_2} \ \mu_2+ \dots +
q_{i_N} \ \mu_N $$
where $i_1,i_2, \dots ,i_N=1,2, \dots N$ and the Weyl orbit $ W(\Lambda^+) $
is formed out of all possible permutations.

One could say that such a lemma is trivial in view of the fact that Weyl
groups of $A_{N-1}$ algebras are nothing but the permutation groups. One must
note however that this can be seen only by using the so-called
fundamental weights defined above.

We now introduce the generators defined by
$$ K_M(u_1,u_2, \dots ,u_N) \equiv u_1^M + u_2^M + \dots + u_N^M
\eqno(II.11) $$
and their generalizations
$$ K_{(q_1,q_2, \dots ,q_N)}(u_1,u_2,..,u_N) \equiv \sum_{j_1,j_2, .. j_N=1}^N
(u_{j_1})^{q_1} (u_{j_2})^{q_2} ...\ (u_{j_N})^{q_N}  \eqno(II.12) $$
for any partition $(q_1,q_2, \dots ,q_N)$ with weight M.
In (II.12), no any two of indices $ j_1,j_2, .. j_N $ shall take the same value
for each particular monomial. For some other reasons, we would like to call
them {\bf class functions}. On the contrary to generic definition of Schur
Functions, the values of M and N need not be correlated in both of these
expressions. One must note here that, as a result of its very definition
given in (II.12), any class function $ K_{(q_1,q_2, \dots ,q_N)} $ has
always a non-linear expansion in terms of generators $ K_M $ defined in
(II.11). We call these expansions {\bf reduction rules}. One can give
$$ \eqalign {
&K_{(q_1,q_2)} = K_{q_1} \ K_{q_2} - K_{q_1+q_2} \ \ , \ \ q_1 > q_2 \cr
&K_{(q_1,q_1)} = {1 \over 2} K_{q_1} \ K_{q_1} - {1 \over 2} K_{q_1+q_1} }
\eqno(II.13) $$
as simple but instructive examples of reduction rules though more
advanced ones can be obtained in exactly the same manner.

It is now seen that (II.10) is equivalent to
$$ ChW(\Lambda^+) = K_{(q_1,q_2, \dots ,q_N)}(u_1,u_2,..,u_N) \eqno(II.14) $$
where we choose the specialization
$$ e(\mu_i) \equiv u_i \ \ , \ \ i=1,2,...N. \eqno(II.15)  $$
Due to (II.3), the indeterminates $u_1,u_2,..,u_N$ here are constrained by
the condition that
$$ u_1 \ u_2 \ ... \ u_N \equiv 1   \eqno(II.16) $$
which can be expressed in the following equivalent form:
$$ K_{(1,1, \dots ,1)}(u_1,u_2,..,u_N) \equiv 1 \ \ .  \eqno(II.17) $$

\hfill\eject

\vskip 3mm
\noindent {\bf{III.\ DEGENERATED SCHUR FUNCTIONS}}
\vskip 3mm

Now we know that, for $A_{N-1}$, there could only be N-1 generic Schur
functions
$$ S_1(x_1) \ , \ S_2(x_1,x_2) \ , \dots \ S_{N-1}(x_1,x_2, \dots x_{N-1}) $$
which can be identified from definition (I.1). For $ M \geq N $, on the other
hand, the Schur functions
$$ S_M(x_1,x_2, \dots ,x_{N-1},x_N, \dots x_M)  $$
transform into some degenerated ones
$$ S_{M \geq N}(x_1,x_2, \dots x_{N-1}) \eqno(III.1)  $$
due to the fact that one can only have N-1 indeterminates. For this,
one must first find an appropriate equivalence between the sets
$ x_1,x_2, \dots ,x_{N-1}$ and $u_1,u_2, \dots ,u_N$ of indeterminates. Hint
comes from completely symmetric functions $h_k(u_1,u_2,..,u_N)$ defined by
$$ \sum_{k=0}^\infty \ h_k(u_1,u_2,..,u_N) \ z^k \equiv
\prod_{i=1}^N \ {1 \over (1-z \ u_i)} \ \ . \eqno(III.2) $$
They can be given equivalently by
$$ h_M(u_1,u_2, \dots ,u_N) =
\sum \ K_{(q_1,q_2, \dots ,q_N)}(u_1,u_2, \dots ,u_N)   \eqno(III.3) $$
where the sum is over all partitions $(q_1,q_2, \dots ,q_N)$ with weight M.
The main observation here is that
$$ h_M(u_1,u_2, \dots ,u_N) \equiv  S_M(x_1,x_2, \dots ,x_M)  \eqno(III.4) $$
is obtained {\bf[1]} by the aid of replacements
$$ K_M(u_1,u_2, \dots u_N) = M \ x_M  \ \ .  \eqno(III.5) $$
From the now on, two sets of indeterminates will always be thought of as the
same via correspondences (III.5). The condition (II.17) can be extended,
on the other hand, as in the following:
$$ K_{(Q,1, \dots ,1)}(u_1,u_2,..,u_N) \equiv K_Q(u_1,u_2,..,u_N)  \ \ , \ \
Q=1,2, \dots  \eqno(III.6) $$
In view of the reduction rules mentioned above, this gives us a way
to introduce extra indeterminates $ x_{Q+N-1} $ which are non-linear
polinomial solutions of (III.6) in terms of indeterminates
$x_1,x_2, \dots ,x_{N-1}$. We have found that the result of all these
calculations leads us to the following formula for the degenerated Schur
functions specified as in (III.1):
$$ S_M = (-1)^N \ S_{M-N-1} - \sum_{k=1}^N \  S_k^{*} \ S_{M-k} \ \ ,
\ \ M \geq N \eqno(III.7) $$
where $S_k^{*}$ is obtained from $S_k$ under replacements
$x_i \rightarrow -x_i $ (i=1,2,..N-1.) . Let us note here that this formula
has a central importance for following applications of our method.

\vskip 3mm
\noindent {\bf{IV.\ EXPLICIT CALCULATION OF WEIGHT MULTIPLICITIES}}
\vskip 3mm

Let
$$ \Lambda^+ \equiv k_1 \mu_1 + k_2 \mu_2  + \dots + k_N \mu_N
\in Sub(M \ \lambda_1)  \eqno(IV.1)  $$
be a dominant weight. In view of (II.14), its character (II.9) will now be
given by
$$ Ch(\Lambda^+) = \sum m_{(q_1,q_2, \dots ,q_N)}(\Lambda^+) \
K_{(q_1,q_2, \dots ,q_N)}(u_1,u_2, \dots ,u_N) \eqno(IV.2)  $$
where
$$ m(\lambda^+ \prec \Lambda^+) \equiv m_{(q_1,q_2, \dots ,q_N)}(\Lambda^+) $$
is assumed for convenience in the notation. The sum in (IV.2) is over all
permutations \ $ (q_1,q_2, \dots ,q_N) $ \ with weight M for which
$$ q_1 \mu_1 + q_2 \mu_2  + \dots + q_N \mu_N
\in Sub(M \ \lambda_1) \ \ . $$
We also know that (IV.1) gives us a Schur function
$$ S_{(k_1,k_2, \dots ,k_N)}(x_1,x_2, \dots ,x_{N-1}) \ \ . $$
Now by stating that
$$ Ch(\Lambda^+) = S_{(k_1,k_2, \dots ,k_N)}(x_1,x_2, \dots ,x_{N-1})
\eqno(IV.3)  $$
one obtains, from the equality of (IV.2) and (IV.3), the
{\bf multiplicity formula}
$$ \sum m_{(q_1,q_2, \dots ,q_N)}(\Lambda^+) \
K_{(q_1,q_2, \dots ,q_N)}(u_1,u_2, \dots ,u_N) =
S_{(k_1,k_2, \dots ,k_N)}(x_1,x_2, \dots ,x_{N-1}) \eqno(IV.4) \ \ . $$
By using the reduction rules mentioned above, this formula turns to an
equation for monomials
$$ x_1^{j_1} \ x_2^{j_2} \dots x_{N-1}^{j_{N-1}}  $$
where $ j_1,j_2, \dots ,j_{N-1}=0,1,2, \dots $ are constrained by
$$ j_1 + j_2 + \dots + j_{N-1} = M-k \ N \geq 0 \ \ ,  \ \  k=0,1,2, \dots
\eqno(IV.5) $$
due to homogeneity. It will depend linearly on multiplicities and the number
of these multiplicities will be equal to the number of monomials given above.
This hence gives us, for any choice of $ \Lambda^+ \in Sub(M \ \lambda_1) $,
the possibility to compute all the multiplicities by solving a system of
linear equation.

Let us visualize our framework, for $A_4$  Lie algebra, in the moderated
example of $ Sub(6 \ \lambda_1) $ which is consist of the following ten
dominant weights:
$$ \eqalign{
&(6,0,0,0,0) \rightarrow 6 \ \lambda_1                         \cr
&(5,1,0,0,0) \rightarrow 4 \ \lambda_1 + \lambda_2             \cr
&(4,2,0,0,0) \rightarrow 2 \ \lambda_1 + 2 \ \lambda_2         \cr
&(3,3,0,0,0) \rightarrow 3 \ \lambda_2                         \cr
&(4,1,1,0,0) \rightarrow 3 \ \lambda_1 + \lambda_3             \cr
&(3,2,1,0,0) \rightarrow \lambda_1 + \lambda_2 + \lambda_3     \cr
&(2,2,2,0,0) \rightarrow 2 \ \lambda_3                         \cr
&(3,1,1,1,0) \rightarrow 2 \ \lambda_1 + \lambda_4             \cr
&(2,2,1,1,0) \rightarrow 2 \ \lambda_1 + \lambda_4             \cr
&(1,0,0,0,0) \rightarrow \lambda_1                               }
\eqno(IV.6) $$
where, due to definitions (II.1) or (II.2), corresponding partitions are given
on the left-hand side. For $A_4$, we have only four Schur functions
$$ S_1(x_1) \ , \ S_2(x_1,x_2) \ , \ S_3(x_1,x_2,x_3) \ , \
S_4(x_1,x_2,x_3,x_4) \ \ . $$
Any other $ S_M(x_1,x_2,x_3,x_4) $ will however be a degenerated Schur
function for $ M \geq 5$ . In this example, we need only
$ S_5(x_1,x_2,x_3,x_4) $ and $ S_6(x_1,x_2,x_3,x_4) $. From definition (I.1),
it is known that
$$ S_5(x_1,x_2,x_3,x_4,x_5) \equiv {1 \over 120} \ x_1^5 +
{1 \over 6} \ x_1^3 \ x_2 + {1 \over 2} \ x_1 \ x_2^2 +
{1 \over 2} \ x_1^2 \ x_3 +  x_2 \ x_3 + x_1 \ x_4 + x_5  \ \ .
\eqno(IV.7)  $$
On the other hand, by solving (II.17), one obtains
$$ x_5 = 1 - {1 \over 120} \ x_1^5 + { 1 \over 6} \ x_1^3 \ x_2 -
{1 \over 2 } \ x_1 \ x_2^2 - {1 \over 2} \ x_1^2 \ x_3 +
 x_2 \ x_3 + x_1 \ x_4 \eqno(IV.8)   $$
(IV.7) and (IV.8) now give us the degenerated Schur function
$$ S_5(x_1,x_2,x_3,x_4) = 1 + {1 \over 3} \ x_1^3 \ x_2 +
2 \ x_2 \ x_3 + 2 \ x_1 \ x_4 \ \ . \eqno(IV.9) $$
In the same manner, one also obtains
$$ \eqalign{
S_6(x_1,x_2,x_3,x_4) = &2 \ x_1 - {1 \over 72} \ x_1^6 +
{1 \over 3} \ x_1^4 \ x_2 - {1 \over 2} \ x_1^2 \ x_2^2 -  \cr
&{2 \over 3} \ x_1^3 \ x_3 + 2 \ x_1 \ x_2 \ x_3 + x_3^2 +
2 \ x_1^2 \ x_4 + 2 \ x_2 \ x_4  } \eqno(IV.10)  $$
from
$$ K_{(2,1,1,1,1)}(u_1, \dots ,u_5)=K_1(u_1, \dots ,u_5) \ \ . $$
One easily sees that both (IV.9) and (IV.10) are compatible with (III.7).
This means that, beside their very definitions as is given above, they can
also be extracted from (III.7) directly.

It is now seen that, from (IV.4), one obtains a linear equation for the
following ten monomials
$$ x_1 \ , \ x_1^6 \ , \ x_1^4 \ x_2 \ , \ x_1^3 \ x_3 \ , \ x_1^2 \ x_2^2 \ ,
\ x_1^2\ x_4 \ , \ x_2^3 \ , \ x_2^2 \ , \ x_2 \ x_4 \ , \ x_1 \ x_2 \ x_3 $$
where their coefficients are linearly depend on the ten multiplicities
corresponding to dominant weights given in (IV.5). Let us choose
$S_{(6,0,0,0,0)}(x_1,x_2,x_3,x_4)$ on the right hand side of (IV.4).
The linear independence of above monomials then gives rise to a result
which is compatible with (II.7). As another example, let us choose
$S_{(5,1,0,0,0)}(x_1,x_2,x_3,x_4)$ on the right hand side of (IV.4).  Then
one has following solutions for corresponding multiplicities:
$$ \eqalign{
&m(6 \ \lambda_1 \prec 4 \ \lambda_1 + \lambda_2)=0                      , \cr
&m(4 \ \lambda_1 + \lambda_2 \prec 4 \ \lambda_1 + \lambda_2)=1          , \cr
&m(2 \lambda_1 + 2 \lambda_2 \prec 4 \ \lambda_1 + \lambda_2)=1          , \cr
&m(3 \ \lambda_2 \prec 4 \ \lambda_1 + \lambda_2)=1                      , \cr
&m(3 \ \lambda_1 + \lambda_3 \prec 4 \ \lambda_1 + \lambda_2)=2          , \cr
&m(\lambda_1 + \lambda_2 + \lambda_3 \prec 4 \ \lambda_1 + \lambda_2)=2  , \cr
&m(2 \ \lambda_3 \prec 4 \ \lambda_1 + \lambda_2)=2                      , \cr
&m(2 \ \lambda_1 + \lambda_4 \prec 4 \ \lambda_1 + \lambda_2)=3          , \cr
&m(\lambda_2 + \lambda_4 \prec 4 \ \lambda_1 + \lambda_2)=3              , \cr
&m(\lambda_1 \prec 4 \ \lambda_1 + \lambda_2)=4  \ \ .    }  $$

\vskip 3mm
\noindent {\bf{V.\ WEYL CHARACTER FORMULA}}
\vskip 3mm

Weyl character formula simply says that
$$ ChR(\Lambda^+) = {A(\rho+\Lambda^+) \over A(\rho)} \ \ \eqno(V.1) $$
where $\rho \equiv \lambda_1 + \lambda_2 + .. + \lambda_{N-1} $ is the
{\bf Weyl vector} of $A_{N-1}$. If one asumes that the left-hand side of
(V.1) is determined by (II.9), then we will show that its right-hand side
gives rise just to the same result hence the formula. The main object is
$$ A(\Lambda^+) \equiv \sum_{\omega} \ \epsilon(\omega) \
e^{\omega(\Lambda^+)} \eqno(V.2) $$
where the sum is over $A_{N-1}$ Weyl group and hence $\omega(\Lambda^+)$
represents an action $\omega$ of the Weyl group. Instead, we will consider
(V.2) always in the following equivalent form:
$$ A(\Lambda^+) \equiv \sum_{\mu \in W(\Lambda^+)} \ \epsilon(\mu) \
e^{\mu} \ \ . \eqno(V.3) $$

\noindent In (V.2), $ \epsilon(\omega) $ is called the {\bf signature} which is
known to be defined by
$$ \epsilon(\omega) = (-1)^{\ell(\omega)} \eqno(V.4) $$
where $\ell(\omega)$ is the minimum number of simple reflections {\bf[3]}
to obtain a Weyl reflection $\omega$. We will however give another definition
of the signature and it is seen that this could be more convenient in
explicit calculations. Having in mind the lemma given in section II, let us
now consider the dominant weight $\Lambda^+$ in the following form
$$ \Lambda^+ = q_1 \mu_1 + \dots + q_s \mu_s  \ \ . \eqno(V.5)  $$
For $ s \leq N $, it is seen that (V.5) is completely equivalent to (II.4).
Having in mind the lemma given
in section II, an element $ \mu$ of its Weyl orbit can be expressed as in
$$ \mu = q_{i_1} \mu_1 + \dots + q_{i_s} \mu_s \eqno(V.6)  $$
where
$$ q_1 \geq q_2 \geq \dots \geq q_s >  0 \ \ .  $$
We can thus replace (V.4) by
$$ \epsilon(\mu) = \epsilon_{q_{i_1},q_{i_2},..,q_{i_s}} \ \ .
\eqno(V.7) $$
The tensor $ \epsilon_{q_{i_1},q_{i_2},..,q_{i_s}} $
is completely antisymmetric in its indices while its numerical value is
given by the condition that
$$ \epsilon(\Lambda^+) \equiv \epsilon_{q_1,q_2,..,q_s} = +1 \ \ . $$

In the specialization (II.15) given above, it is now seen that
$$ A(\rho) = \prod_{j>i=1}^N \ (u_i-u_j)  \eqno(V.8) $$
which is in fact nothing but the Vandermonde determinant. It is also known
that $ A(\rho+\Lambda^+)$ is equivalent to determinant of the following
$N \times N$ matrix:
$$ \pmatrix{
u_1^{q_1-1+N}&u_1^{q_2-2+N}&\ldots&u_1^{q_N-N+N}\cr
u_2^{q_1-1+N}&u_2^{q_2-2+N}&\ldots&u_2^{q_N-N+N}\cr
\vdots&\vdots&\vdots&\vdots&\cr
u_N^{q_1-1+N}&u_N^{q_2-2+N}&\ldots&u_N^{q_N-N+N}\cr} \ \ . \eqno(V.9)   $$
As the main result of this section, one can show that (V.8) and (V.9)
give
$$ A(\rho+\Lambda^+) = A(\rho) \ S_{(q_1,q_2, \dots ,q_N)}(x_1,x_2,..,x_{N-1})
\eqno(V.10)  $$
for the dominant weight $\Lambda^+ = q_1 \mu_1 + \dots + q_N \mu_N$.
A nice example can be given in the case of $\Lambda^+ = M \ \lambda_1$ by
$$ A(\rho+M \ \lambda_1) = A(\rho) \ S_M(x_1,x_2,..,x_{N-1})  \eqno(V.11) $$
which is nothing but a simple result of (II.7). Note here that, for $M \geq N$,
(V.11) gives us always a degenerated schur function. For any other
$\Lambda^+ \in Sub(M \ \lambda_1)$, (V.10) gives on the right hand side the
Schur functions which can be obtained via (I.3). In result, this shows us
that our statement (IV.3) is in complete agreement with Weyl character formula.

\vskip3mm
\noindent{\bf {REFERENCES}}
\vskip3mm

\item[1] V.G.Kac and A.K.Raina, Bombay Lectures on Highest Weight Representations,
\ \ \ \ \ World Sci., Singapore (1987)

\item[2] H.Freudenthal, Indag. Math. 16 (1964) 369-376
\item
\ \  G.Racah, Group Theoretical Concepts and Methods in Elementary Particle Physics, ed. F.Gursey, N.Y.,
Gordon and Breach (1964)
\item
\ \      B.Kostant, Trans.Am.Math.Soc., 93 (1959) 53-73

\item[3] J.E.Humphreys, Introduction to Lie Algebras and Representation Theory, N.Y., Springer-Verlag (1972)

\item[4] V.G.Kac, Infinite Dimensional Lie Algebras, N.Y., Cambridge Univ. Press (1990)

\item[5] H.R.Karadayi, Anatomy of Grand Unifying Groups I and II,
\item \ ICTP preprints IC/81/213 and 224
\item \ \  H.R.Karadayi and M.Gungormez, Jour.Math.Phys., 38 (1997) 5991-6007

\end